\documentclass[
mathleft
]{an}
\usepackage{graphicx}
\usepackage{times}
\begin{document}

\sloppy
\setlength{\mathindent}{0pt} 

% The following seven commands are intended for editorial usage and
% should be ignored by the author(s).
\Pagespan{1}{}% Document's page range.
% If second parameter is left empty, the last page is computed
% automatically.
\Yearpublication{XXX}%
\Yearsubmission{2010}%
\Month{6}%
%\Volume{}%
%\Issue{}%
% \DOI{}%

\title{Secular variation of hemispheric phase differences in the solar cycle}

\author{N.V.~Zolotova\inst{1}\fnmsep\thanks{Corresponding author:
  {ned@geo.phys.spbu.ru}}
\and  D.I.~Ponyavin\inst{1}\and
R.~Arlt\inst{2}
\and  I.~Tuominen\inst{3}
}
\titlerunning{Secular variation of hemispheric phase differences in the solar cycle}
\authorrunning{N.V.~Zolotova et al.}
\institute{
Institute of Physics, St. Petersburg State University, Ulyanovskaya yl. 1, Petrodvorets, St. Petersburg, 198504, Russia
\and
Astrophysikalisches Institut Potsdam, An der Sternwarte 16,
D-14482 Potsdam, Germany
\and
Observatory, P.O. Box 64
FI-00014, University of Helsinki, Finland}

\received{2010 Jun 14}
\accepted{XXX}
\publonline{XXX}


\abstract{%
   We investigate the phase difference of the sunspot cycles in the two hemispheres and compare it with the latitudinal sunspot distribution. If the north-south phase difference exhibits a long-term tendency, it should not be regarded as a stochastic phenomenon.
   We use datasets of historical sunspot records and drawings made by Staudacher, Hamilton, Gimingham, Carrington, Sp\"{o}rer, and Greenwich observers, as well as the sunspot activity during the Maunder minimum reconstructed by Ribes and Nesme-Ribes.
   We employ cross-recurrence plots to analyse north-south phase differences. We show that during the last 300 years, the persistence of phase-leading in one of the hemispheres exhibits a secular variation. Changes from one hemisphere to the other leading in phase were registered near 1928 and 1968 as well as two historical ones near 1783 and 1875.
   A long-term anticorrelation between the hemispheric phase differences in the sunspot cycles and the latitudinal distribution of sunspots was traced since 1750.}

\maketitle

%________________________________________________________________

\section{Introduction}
%________________________________________________________________

In order to deal with the asymmetry of hemispheric sunspot distributions, separate observations for the northern and southern hemispherical activities are required. Suitable monitoring was started by the Royal Greenwich Observatory in 1874. Now daily information about size, position and area of sunspots are comprised by the RGO/USAF/NOAA dataset. Unfortunately, both the International Sunspot Numbers and the Group Sunspot Numbers are defined for the entire visible solar disk only.

Applying the cross-recurrence plot technique to the smoothed sunspot area data, we compared the time-varying phase differences between the northern and southern hemispheres with the latitudinal distribution of the sunspots (Zolotova et al.~\cite{zolotova_2009}).
If the phase difference between the northern and southern hemispheres is positive, we refer to the northern hemisphere as `leading in phase'. If the difference is negative, the southern hemisphere is `leading in phase'. We observed a long-term persistence of one hemisphere exhibiting a leading phase, which lasts almost four solar cycles, while after that, the opposite hemisphere leads in phase. The full period corresponds to the secular cycle first described by Gleissberg (\cite{gleissberg_1967}). We also verified that these long-term variations in the hemispheric phase differences are in anticorrelation with the long-term variations of the latitudinal distribution of the sunspots -- {\it the magnetic equator} (Pulkkinen et al.~\cite{pulkkinen_1999}).

Since the analysed time series starts in May 1874, we were unable to detect the change in sign of the phase difference before the 12th cycle. In the present paper we extend our previous studies of historical hemispheric sunspot asynchrony as far back as to the Maunder minimum.
Direct observations of the Sun by Staudacher (Arlt~\cite{arlt_2008}, \cite{arlt_2009a}), Hamilton \& Gimingham (Arlt~\cite{arlt_2009b}), Carrington (\cite{carrington_1863}), and Sp\"{o}rer (\cite{sporer_1874}, \cite{sporer_1878}, \cite{sporer_1880}, \cite{sporer_1886}, \cite{sporer_1894}) are involved. Further, we reconstruct and discuss the temporal variation of the phase difference and the latitudinal distribution of the sunspots during the last 300 years.

%________________________________________________________________

\section{Data and methods}
%________________________________________________________________

For the analysis we use the following list of daily data:
\begin{itemize}

\item[--] The drawing of the butterfly diagram during the Maunder
    minimum constructed by Ribes \& Nesme-Ribes~(\cite{ribes_1993}).

\item[--] The sunspot drawings by J.C.~Staudacher for the period 1749--1796. Their digitization, determination of heliographic latitudes and longitudes in the Carrington rotation frame, and reconstruction of the butterfly diagram was done by Arlt (\cite{arlt_2008}, \cite{arlt_2009a}). These data were extended with a few solar observations by J.A.~Hamilton and W.~Gimingham for the period 1795--1797 (Arlt~\cite{arlt_2009b}).

\item[--] The records by R.C.~Carrington for the period 1853--1861. His famous book contains the heliographic coordinates of individual sunspots (Carrington \cite{carrington_1863}). The last part of the book represents sunspot group drawings for each rotation, starting from the first in 9 November 1853. One of Carrington's books was presented by the Royal Astronomical Society to the Russian Imperial Library (now the Library of the Russian Academy of Science), and one to the Helsinki University library.

\item[--] The records of heliographic sunspot positions by G.~Sp\"{o}rer for the period 1861--1894 (Sp\"{o}rer \cite{sporer_1874}, \cite{sporer_1878}, \cite{sporer_1880}, \cite{sporer_1886}, \cite{sporer_1894}). All books consist of a description, tables with positions, and drawings of sunspot groups at their latitudes; the longitude is not shown in the drawings, instead, the horizontal axis is used for the date. These observations are not ordered by calendar dates, but by a sequence of group numbers. This data and Carrington's were digitized in the Helsinki Observatory about 20--30 years ago and arranged as a function of time.

    \item[--] The RGO/USAF/NOAA dataset from 1874 to the present, originally in the Royal Observatory,
 Greenwich~\cite{rgo_1874_1976}).
    %In contrast to the above historical observations, the Greenwich data do not contain days without observations (gaps).
\end{itemize}

Finally, we compiled a list of heliographic positions of sunspots for those days, when observations occur (except for the Maunder minimum, when only the butterfly diagram is available; Ribes \& Nesme-Ribes~\cite{ribes_1993}).
%--- Fig.~\ref{Figure3})   [left out to keep the order of figures - RA]
To count sunspots for each hemisphere we replaced the latitude values by plus ($+1$) and minus ($-1$) respectively for the northern and southern hemispheres. Next, we added numbers belonging to the same day but separately for the hemispheres.

To measure the phases differences of sunspot cycles in the two hemispheres we have used the {\it Cross-Recurrence Plot} (CRP) method (Marwan et al.~\cite{marwan_2007}), where two time series are embedded in the same phase space. The CRP analyses the parallel occurrence of states as described by the time series $x_i$ and $y_i$ and is defined by the cross-recurrence matrix $\mathbf{CR}$:

\begin{equation}
{\rm CR}_{i,j} = \Theta(\varepsilon_{i}-\|x_{i} - y_{j}\|),
\qquad i,\,j = 1,\ldots, N,
\end{equation}
where $N$ is the length of the time series $x$ and $y$, $\|\cdot\|$ is a norm, and $\Theta$ is the Heaviside function (Marwan et al.~\cite{marwan_2007}). The adjustable parameter $\varepsilon_i$ is called the recurrence point density.

In the special case of $x = y$, the main diagonal { CR}$_{i,i}$ consists of
ones ({ CR}$_{i,i}$ = 1, $i = 1,\, \ldots\, ,N$). For this reason it is called the
{\it line of identity}. Applying a transformation of the timescale (e.g.,~a
phase shift) to the second time series $y$, the line of identity becomes bowed
or shifted (Marwan et al.~\cite{marwan_2002}; Marwan \& Kurths \cite{marwan_2005}). It is called the {\it line of synchronisation} (LOS), because one can use it to rescale the timescale of the second time series and determine the closest match (synchronisation) of the two time series. %We, therefore, note that this synchronisation has a different meaning than the phase synchronisation as a physical phenomenon.

The extraction of the LOS with the CRP is performed by the algorithm proposed in Marwan et al.~(\cite{marwan_2002}), available in the CRP Toolbox for Matlab\textsuperscript{\textregistered}. When plotting the matrix with $i$ and $j$ being the axes of a diagram (see Sect.~\ref{results}), the LOS can be understood as a construction from the lower left to the upper right corner of the CRP-pattern, using the distribution of recurrence points within a window of size ${dx\! \times\! dy}$ with its origin at the identified recurrence point $(i,j)$.

%
%________________________________________________________________

\section{Results\label{results}}

%________________________________________________________________

\begin{figure*}
    \centering
 \includegraphics[width=17cm]{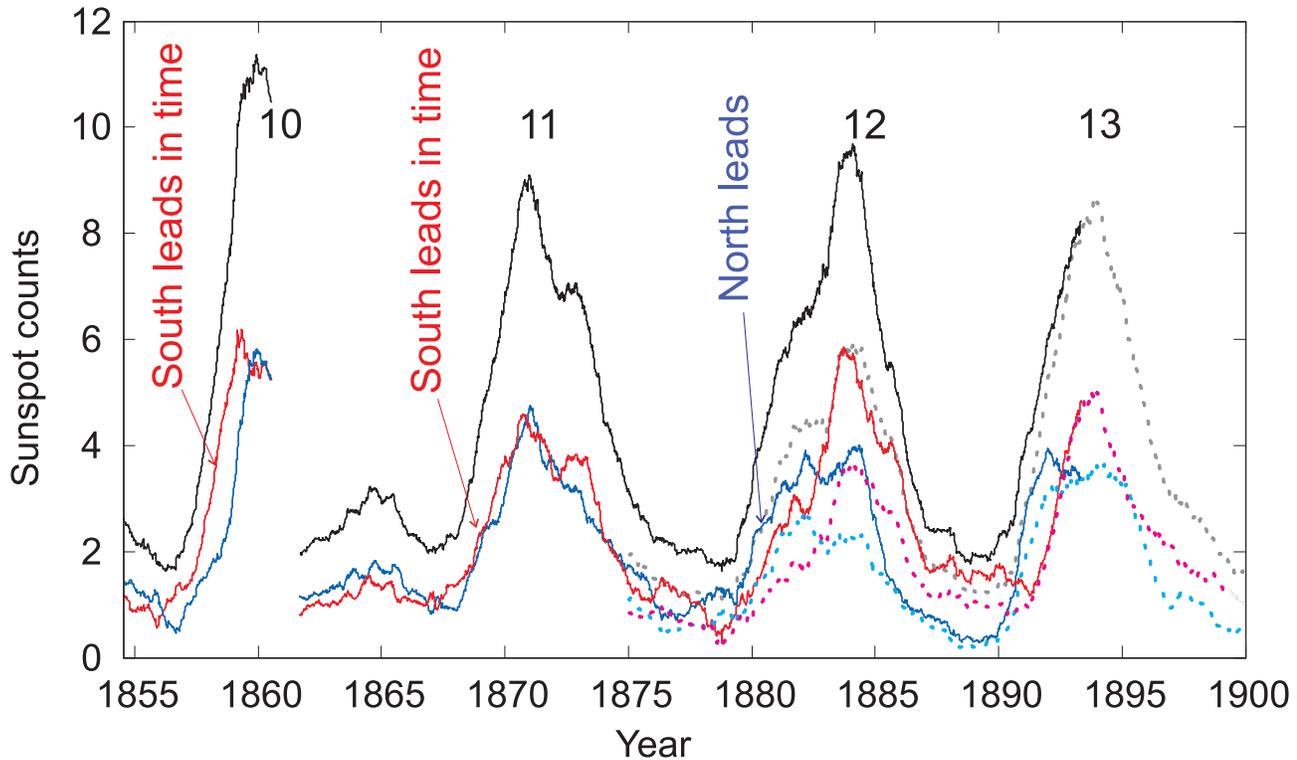}
    \caption{ Smoothed daily sunspot counts according to the
    observations by Carrington and Sp\"{o}rer for the north (blue), south (red), and
    the entire solar disk (black). The Greenwich data are marked by dashed lines in light colours
    (light-blue, pink and grey correspondingly). The cycle numbers refer to the Z\"{u}rich numbering. }
              \label{Figure1}
    \end{figure*}

From the analysis of the Greenwich sunspot area data it was shown that the consecutive sign changes in the phase difference between the two hemispheres occurred at the maximum of the 16th cycle and then at the minimum between the cycles 19 and 20 (Zolotova et al.~\cite{zolotova_2009}). At the same times the latitudinal distribution of the sunspots (expressed by the magnetic equator, defined by the sunspot latitudes) changed its sign (Pulkkinen et al.~\cite{pulkkinen_1999}). It is notable that the phase difference between the northern and southern hemispheres is in anticorrelation with the latitude of the magnetic
equator.

The raw daily sunspot counts reconstructed from the observations by Carrington, Sp\"{o}rer, Staudacher, Hamilton, and Gimingham are not as uniform as the Greenwich data. Staudacher made his drawings at rather irregular time intervals. Only sometimes, he carried out more that 20--30~observations during a year. Carrington and Sp\"{o}rer usually did not admit gaps longer than 10~days, but they did not report day-by-day observations either.

Averaging a time series with observational gaps is a commonly encountered problem when historical solar data are used (Vaquero~\cite{vaquero_2007}; Usoskin~\cite{usoskin_2003}). Since raw daily sunspot counts vary strongly, but we are interested in quite smooth curves, we used simple sliding averages over 500~days (for the Carrington and Sp\"{o}rer data) and over 1000~days (for the Staudacher, Hamilton, and Gimingham data). Smoothing with such wide windows influences the cycle minimum and maximum levels, and causes time shifts. Thus, a strict numerical estimation of the amplitude of the northern and southern hemispheres or absolute phases may be incorrect. However, we are interested in phase differences and the hemispheres' behaviour during their ascending and descending phases of the solar cycle evolution (Zolotova et al.~\cite{zolotova_2009}).

     \begin{figure}
    \centering
 \includegraphics[width=8.3cm]{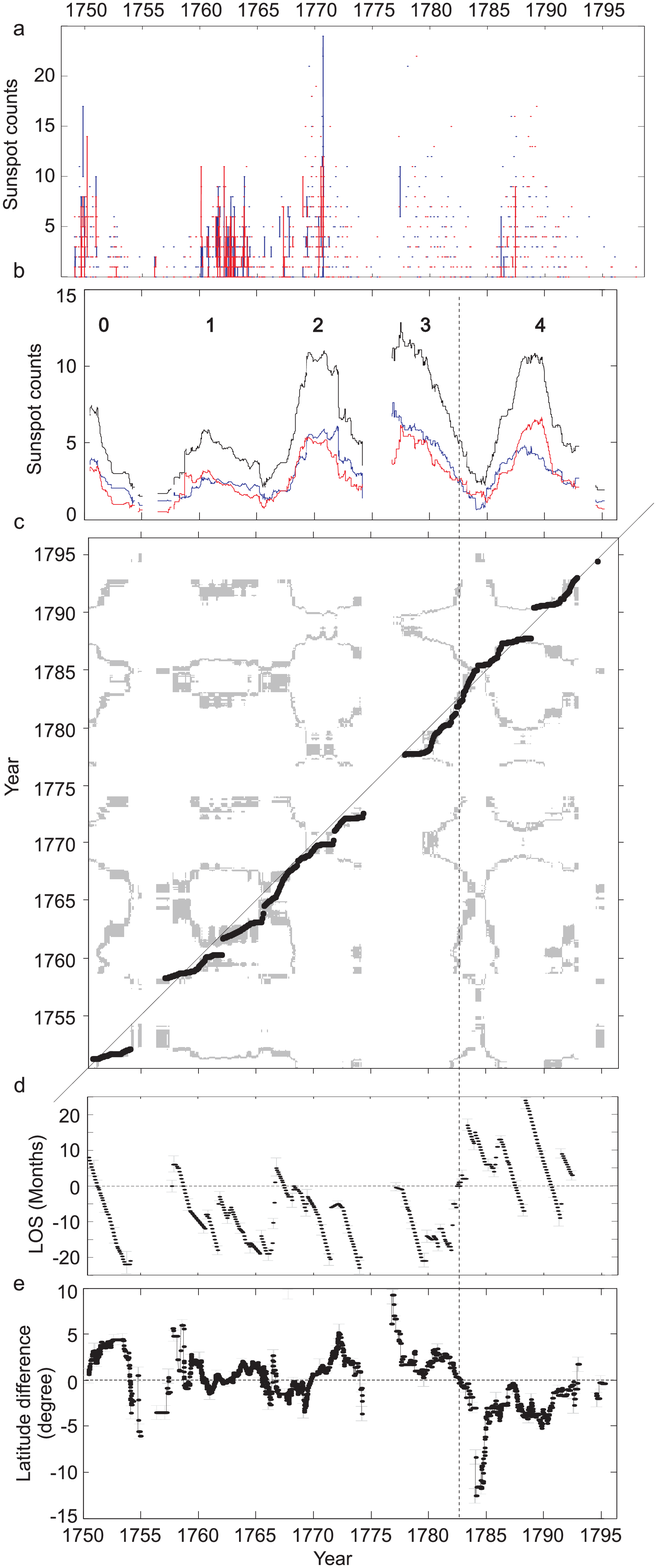}
    \caption{ \emph{a}) Raw daily sunspot counts according to the observations
    by Staudacher, Hamilton, and Gimingham for the northern (blue) and southern
    (red) hemispheres. \emph{b}) Smoothed sunspot counts based on the same data set
    for the north (blue),  south (red), and the entire solar disk (black).
    The cycle numbers refer to the Z\"{u}rich numbering. \emph{c})
Cross-recurrence plot of the resampled, smoothed sunspot counts. The line of synchronization (LOS) is
presented as black thick line. \emph{d}) Extracted LOS. \emph{e}) Latitude difference of sunspots distribution (the magnetic equator) versus time variations. Error bars indicated by gray colour.
    The vertical dashed line denotes the change of leading near 1783\,$\pm$\,1.5 yr.}
              \label{Figure2}
    \end{figure}

Smoothed sunspot counts reconstructed from the observations by Carrington and Sp\"{o}rer for the north (blue), south (red), and the entire solar disk (black) are shown in Fig.~\ref{Figure1}. The smoothing procedure shortens the time-series by half of the window length at the beginning and the end of the raw data. Therefore, the smoothed Carrington and Sp\"{o}rer sunspot counts do not overlap (Fig.~\ref{Figure1}), although Carrington finished observing by the end of March 1861 and Sp\"{o}rer started in January 1861.
 % Spoerer started his astronomical obs. in Anklam in 1860
 % in general. So, 1861 is a likely start for the solar obs.
 % He did not work on sunspots when he was still in Berlin
 % before. [RA]
During these about three months Carrington registered systematically more sunspots in comparison to Sp\"{o}rer. We suggest that the discrepancies can be explained by a difference in detecting small sunspots or pores on the solar disc.

The Sp\"{o}rer and Greenwich data overlap during a period of about 19.5~years. Both data sets were averaged with the same 500-day window. The Greenwich data are given by dashed lines in Fig.~\ref{Figure1}. During the maximum of the 12th cycle, Sp\"{o}rer registered on average two more sunspots in each hemisphere than the Greenwich observers. The next cycle does not show this difference. Nevertheless, both Sp\"{o}rer and Greenwich datasets show that from the beginning of cycle 12, the northern hemisphere leads in time with respect to the southern. During the ascending phase of cycle 11, there is no hemisphere clearly leading, but the southern hemisphere slightly predominates in leading continuously. The ascending phase of cycle 10 also shows a leading southern hemisphere. Thus, we can claim the change of southern phase-leading into northern between the maximum of cycle 11 and the following minimum. We also confirm an earlier finding (Pulkkinen et al.~\cite{pulkkinen_1999}; Zolotova et al.~\cite{zolotova_2009}) that the magnetic equator lies in the hemisphere which is not leading in phase.

In Fig.~\ref{Figure2}a raw sunspot counts reconstructed from the observations by Staudacher, Hamilton, and Gimingham are shown for the northern (blue) and southern (red) hemispheres. Day-to-day observations (without gaps) are joined by lines. Some periods (solar maxima) are covered by frequent observations, but there are several gaps longer than half a year during solar minima. The average was not calculated if within the 1000-day window more than 990~days were without observations. Such periods correspond to the minima before the cycles~1 and~3, and the descending phase of the 4th cycle (Fig.~\ref{Figure2}b, the entire solar disk sunspot counts are shown in black colour) (Arlt~\cite{arlt_2008}, \cite{arlt_2009a}).

Similarly to Zolotova \& Ponyavin~(\cite{zolotova_2006}) and Zolotova et al.~(\cite{zolotova_2009}), to study the time-varying phase difference of the northern and southern hemispheres, we used the cross-recurrence plot (CRP) technique~(Marwan et al.~\cite{marwan_2007}). The algorithm is available in the CRP Toolbox for Matlab\textsuperscript{\textregistered}.

Since the original Staudacher, Hamilton, and Gimingham records are too long for the CRP analysis, the data were resampled\footnote{Matlab (and the CRP toolbox as its application) operates slowly with loops for big matrices. The cross-recurrence plot is an $N\times N$ matrix (where $N$ is the time-series length). When $N$ is more then some value (depends on the processor and RAM of the computer) Matlab toolbox stops operation or produces errors. A personal computer is able to process data with $N$ equal to about 3500--5000. For the non-resampled Staudacher, Hamilton and Gimingham data, $N=16826$. Fortunately, the plots of non-resampled and resampled data are highly similar.}. We extracted each 30th value from the smoothed daily sunspot counts. In a simplified attempt to construct the CRP which is sufficient for our purposes, these resampled numbers go into the time series $x_i$ and $y_i$. Short resampled time series are practically identical to non-resampled, smoothed data shown in Fig.~\ref{Figure2}b. Figure~\ref{Figure2}c shows the cross-recurrence plot (gray pattern) of the resampled, smoothed  sunspot counts with a diagonal line for orientation. The CRP was calculated using a fixed recurrence point density RR, $\varepsilon=0.1$. For the construction of LOS, the parameters $dx=30$ and $dy=30$ were used.

LOS detects differences in timescales between two data series or dynamical systems (black thick line in Fig.~\ref{Figure2}c). LOS variations relative to the zero level (or line of identity) are shown in Fig.~\ref{Figure2}d. The change in sign of LOS indicates the change of phase-leading from one hemisphere to the other (LOS $>$ 0 -- north precedes south, LOS $<$~0 -- south precedes north). During the period from near 1750 to 1783\,$\pm$\,1.5 yr (the uncertainty comes from half of the averaging window), the southern hemisphere predominantly leads. Afterwards, the entire 4th cycle is phase-led by the northern hemisphere. To account for the effect of finite statistics, we used a crude approximation of errors $\sim 1/C^{1/2}$, where $C$ is the sunspot counts on the entire visible solar disk during a day.

Next, the magnetic equator was calculated as the sum of the mean latitudes $\langle \lambda\rangle$ of sunspots in the northern and southern hemispheres $\langle \lambda(N)\rangle _{n} + \langle \lambda(S)\rangle _{n}$ (southern latitudes being negative), where $n$ is the time window, and is shown in Fig.~\ref{Figure2}e. The window size is the same -- 1000~days, the requirement on free days from observations is also the same -- no more than 990 out of 1000~days. The uncertainty in the sunspot latitude definition does not exceed $15^{\circ}$ (Arlt~\cite{arlt_2009a}). It is seen that from about 1750 to 1783 the magnetic equator was predominantly located in the northern hemisphere, while after that, in the beginning of cycle 4, it rapidly moved to the southern hemisphere. This again is in precise anticorrelation with the hemisphere leading in phase (vertical dashed line).

Recently it was suggested that the unusual length of the 4th solar cycle can be explained by an outstanding phase asynchrony between the northern and southern hemispheric activities reaching a delay of up to 4.5 years during the course of its ascending phase (Zolotova \& Ponyavin~\cite{zolotova_2007}). However, from Fig.~\ref{Figure2}d it is seen that the delay, during the first half of the cycle 4, between north and south does not exceed two years. The hypothesis about the lost cycle 4 just before the Dalton minimum is actually discussed by Usoskin et al.~(\cite{usoskin_2009}).

  \begin{figure*}
    \centering
 \includegraphics[width=16cm]{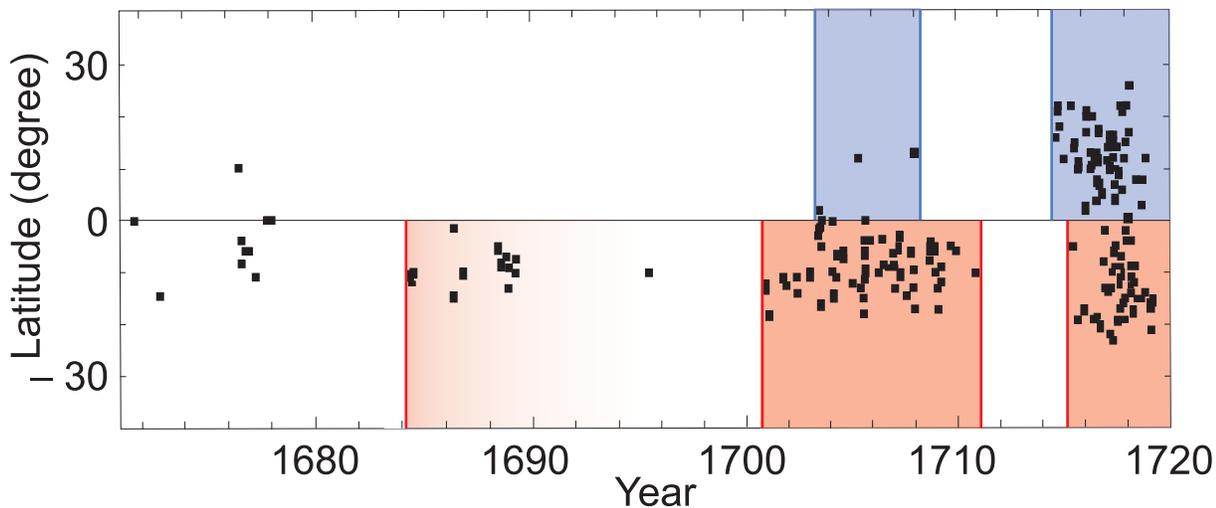}
    \caption{ Butterfly diagram of sunspots
reconstructed for the period 1670--1719 according to Ribes \&
Nesme-Ribes (\cite{ribes_1993}). Vertical coloured lines mark
activity onsets and ends in hemispheres: north -- blue, south -- red.}
              \label{Figure3}%
    \end{figure*}

During the Maunder minimum (1645--1715) the butterfly symmetry was broken, and a significant amplitude difference between the hemispheres became evident. Sunspots were commonly observed in the southern hemisphere (Ribes \& Nesme-Ribes~\cite{ribes_1993}; Sokoloff \& Nesme-Ribes~\cite{sokoloff_1994}). But what about the phase difference of hemispheres during that epoch?

Figure~\ref{Figure3} is a copy of the butterfly diagram from Ribes
\& Nesme-Ribes~(\cite{ribes_1993}). Vertical blue lines mark
activity onsets and ends in the northern hemisphere, red lines --
in the southern hemisphere. From 1683 to 1718 Philippe
de la Hire conducted solar observations at the Observatoire de Paris.
He carried out at least 10~observations (usually more) in each month.

The north-south phase differences before the 1680s are difficult to fix, because sunspots lie very near the equator. In addition to that, the observations, before the 1680s, were irregular and rare (Ribes \& Nesme-Ribes~\cite{ribes_1993}).

Near 1701 the activity in the southern hemisphere starts about two years earlier than in the opposite one. The next solar cycle starts near 1715 and the north leads the south by about one year. Thus, the sign change in the hemispheric phase difference probably occurred between 1706 and 1715. However, up to 1714 the northern activity was very weak. Hence, our hypothesis of the change of phase-leading from south to north should be considered a tentative suggestion. At the same time, up to 1714 the latitudinal distribution of sunspots was evidently shifted to the southern hemisphere.

After the Maunder Minimum during the period from 1715 to 1719 when the solar cycle recovered, the magnetic equator cannot be determined easily simply by the eye. However, the northern activity precedes the southern one by about one year.

%
%________________________________________________________________

\section{Discussion}

%________________________________________________________________

  \begin{figure*}
    \centering
 \includegraphics[width=16cm]{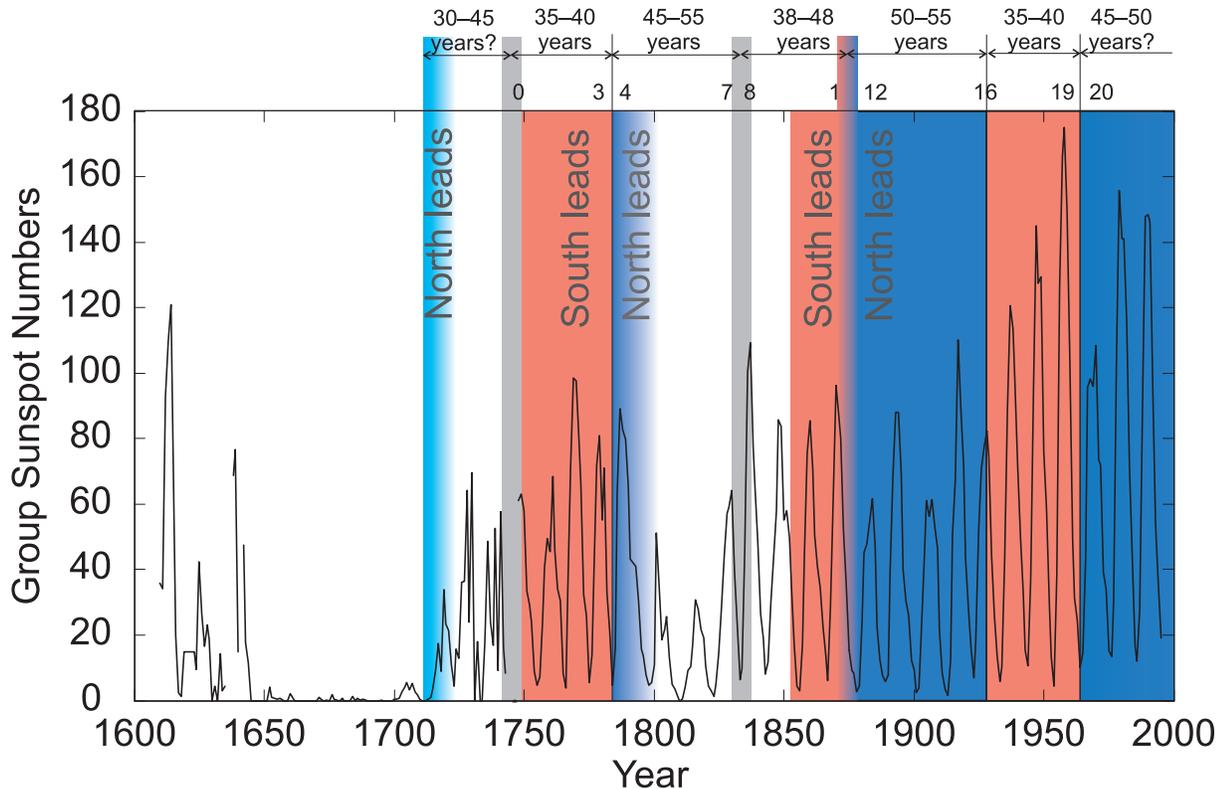}
    \caption{ Yearly group sunspot numbers versus time. Periods of the
    northern hemisphere preceding the southern one are marked by blue
    colour, periods when the southern hemisphere precedes the northern
    one are marked by red colour.
Vertical lines denote the sign changes of the phase difference. Grey
bars indicate probable sign changes. The cycle numbers refer to the
 Z\"{u}rich numbering.}
              \label{Figure4}%
    \end{figure*}

 We investigate variations of the hemispheric phase difference in the solar cycle in the  period of 1670--2009 by a combination of positional sunspot data  from various sources. The results are shown in Fig.~\ref{Figure4} together  with the yearly group sunspot numbers by Hoyt \& Schatten (\cite{hoyt_1998}) for 1610--1995.

We can summarize that during 1715--1719 the northern-hemisphere cycle precedes the southern one (Fig.~\ref{Figure4}, corresponding to the time interval coloured by light-blue). Since this conclusion was made only from a visual analysis of the butterfly diagram during the Maunder minimum, it is not a robust result. The northern-hemisphere leading probably lasts 30--45~years until a sign change in the phase difference occurs near 1745--1750.
Then, according to the Staudaucher's drawings, the southern hemisphere precedes for 35--40~years and hands over to the northern hemisphere near 1783. The next recurrent change in phase-leading is supposed to lie somewhere between the maxima of the 7th and 8th cycles (grey bar near 1833). This change in phase-leading is not documented by data yet. Such a long interval for a sign change (roughly 10~years) is chosen because the secular variations of phase asymmetry as well as of the magnetic equator are not symmetric sinusoidal functions. An individual phase-leading period by either hemisphere may last somewhat longer or somewhat shorter than 4~cycles. Thus, the northern hemisphere preceded the southern one from the beginning of the cycle~4, probably for 45--55 years. Accordingly, the southern hemisphere then preceded the northern one for 38--48 years.

According to the Carrington and Sp\"{o}rer data, the southern hemisphere leads in phase during the 10th cycle and during the ascending phase of the 11th cycle, but by the onset of the 12th cycle the northern hemisphere starts to precede the southern one. This sign-change in phase difference happened gradually, during the descending part of the 11th cycle. Note that, in contrast, another north-south change in phase-leading occurring in the minimum between cycles 19 and 20 was abrupt (Zolotova et al.~\cite{zolotova_2009}).

Further long-term variations
in the hemispheric phase differences are described by Zolotova et al.~(\cite{zolotova_2009}): from the beginning of the 12th to the maximum of the 16th cycle, the northern hemisphere dominates in leading for about 50--55 years period, after which, until the minimum before the cycle 20, the southern hemisphere leads for about 35--40 years, and then, till the present, in spite of depressed sunspot activity during the new solar cycle 24, the northern hemisphere leads in time. To keep pace with the proposed secular variation in hemispheric leading, we expect the next sign change in phase difference as well as the northward shift of the magnetic equator will happen no later than in the end of the current solar cycle (2018--2020). For convenience, the results are also shown in Table~1. The hemispheres are not equal in their duration of phase-leading. The duration of the northern hemisphere preceding the southern one is longer than the opposite case. The cause of such a skewness is unclear.

\begin{table}
\caption{Preceding hemispheres over the last 300~years.} % title of Table
\label{table:1} % is used to refer this table in the text
%\centering % used for centering table
\begin{tabular}{l c c c} % centered columns (4 columns)
\hline\noalign{\smallskip}% inserts double horizontal lines
Preceding & Onset  & End  & Duration \\ % table heading
Hemisphere &  &  & [yr] \\[1.5pt]
\hline\noalign{\smallskip} % inserts single horizontal line
%Southern & --- & 1706--1715 & --- \\ % inserting body of the table
Northern & 1706--1715 & 1745--1750 & 30--45 \\
Southern & 1745--1750 & 1783 & 35--40\\
Northern & 1783  & 1829--1837 & 45--55 \\
Southern & 1829--1837 & 1873--1877 & 38--48 \\
Northern & 1873--1877 & 1928  & 50--55 \\
Southern & 1928    & 1968  & 35--40 \\
Northern & 1968 & --- & --- \\[1.5pt]
\hline %inserts single line
\end{tabular}
\end{table}

From 1750 to 1796 and from 1853 to the present, phase-leading is found to be anticorrelated with the latitudinal distribution of the sunspots. That is, the magnetic equator will be located in the hemisphere which lags behind the other hemisphere in its activity. However, such relationship is not continuous, and was broken during the Maunder minimum. Probably it can be explained by qualitatively different dynamo behaviour during Grand minima compared with the normal dynamo-states (Brandenburg \& Spiegel~\cite{brandenburg_2008}; Brooke et al.~\cite{brooke_1998}; K\"{u}ker et al.~\cite{kuker_1999}; Ossendrijver~\cite{ossendrijver_2003}).

Note that, interpreting the solar cycle as a pair of activity waves travelling towards the equator and pole respectively, Pelt et al.~(\cite{pelt_2000}) found the Gleissberg cycle modulation not only in the variations of the magnetic equator, but in the period length of the Schwabe cycle as well.

\newpage

%__________________________________________________________________

\section{Conclusion}

%__________________________________________________________________

We used historical documents by Staudacher, Hamilton, Gimingham, Carrington, Sp\"{o}rer, and the Greenwich observers, as well as the sunspot activity reconstruction by Ribes \& Nesme-Ribes (\cite{ribes_1993}). We showed that one of the two hemispheres is always slightly advanced in the progress of its cyclic activity compared to the other hemisphere during the last 300~years. After several cycles, the hemispheres change their roles. The long-term persistence of this phase-leading does not demonstrate a clear periodicity, but varies in duration around 8~solar cycles. The periodicity probably corresponds to the Gleissberg cycle. Phase-leading is found to be anticorrelated with the latitudinal distribution of the sunspots. This means that the sunspots in a preceding hemisphere show a butterfly wing emerging on a slightly lower latitudes as compared to the butterfly wing of the opposite, delayed hemisphere, where the sunspots emerged at slightly higher latitudes. Since the north-south asymmetry exhibits a long-term persistence, it should not be regarded as a stochastic phenomenon.

%__________________________________________________________________

%__________________________________________________________________

\acknowledgements

For the calculation of CRP and LOS, we have used the CRP Toolbox, developed by Norbert Marwan,
for Matlab\textsuperscript{\textregistered}, available at
tocsy.pik-potsdam.de.
The Royal Greenwich Observatory USAF/NOAA daily data of sunspot
area and their latitudes separately for the northern and southern
hemispheres is available at
http://solarscience.msfc.nasa.gov/greenwch.shtml.
The group sunspot numbers (GSN) data provided by Hoyt \& Schatten
(\cite{hoyt_1998}) and tabulated by the National Geophysical Data
Center is available at
http://www.ngdc.noaa.gov/stp/SOLAR/ ftpsunspotnumber.html.


\begin{thebibliography}{}
\bibliographystyle{an}

\bibitem[2008]{arlt_2008} Arlt, R.: 2008, Sol. Phys. 247, 399

\bibitem[2009a]{arlt_2009a} Arlt, R.: 2009a, Sol. Phys.  255, 143

\bibitem[2009b]{arlt_2009b} Arlt, R.: 2009b, AN 330, 311

\bibitem[2008]{brandenburg_2008} Brandenburg, A., Spiegel, E.A.: 2008, AN 329, 351

\bibitem[1998]{brooke_1998} Brooke, J.M., Pelt, J., Tavakol, R., Tworkowski, A.:  1998, ApJ 332, 339

\bibitem[1863]{carrington_1863} Carrington, R.C.: 1863, {\it Observations of the Spots on the Sun from
November 9, 1853 to March 24, 1861 made at Redhill}, Williams and Norgate, London, Edinburgh

\bibitem[1967]{gleissberg_1967} Gleissberg, W.: 1967, Sol. Phys. 2, 231

\bibitem[1998]{hoyt_1998} Hoyt, D.V., Schatten, K.: 1998, Sol. Phys. 179, 189

\bibitem[1999]{kuker_1999} K\"{u}ker, M., Arlt, R., R\"{u}diger, G.: 1999, A\&A 343, 977

\bibitem[2005]{marwan_2005} Marwan, N., Kurths, J.: 2005, Phys. Lett. A 336, 349
\bibitem[2002]{marwan_2002} Marwan N., Thiel, M., Nowaczyk, N.R.: 2002, Nonlin.~Proc.~Geophys. 9, 325



\bibitem[2007]{marwan_2007} Marwan, N., Romano, M.C., Thiel, M., Kurths, J.: 2007, Phys.~Rep. 438, 237

\bibitem[2003]{ossendrijver_2003} Ossendrijver, M.: 2003, A\&A Rev. 11, 287

\bibitem[2000]{pelt_2000} Pelt, J.,  Brooke, J., Pulkkinen, P.J., Tuominen, I.: 2000, A\&A 362, 1143

\bibitem[1999]{pulkkinen_1999} Pulkkinen, P.J., Brooke, J., Pelt, J., Tuominen, I.: 1999, A\&A 341, L43

\bibitem[1993]{ribes_1993} Ribes, J.C., Nesme-Ribes, E.: 1993, A\&A 276, 549

\bibitem[1874--1976]{rgo_1874_1976} Royal Observatory, Greenwich: 1874--1976, Greenwich Photoheliographic Results, in 103 volumes

\bibitem[1994]{sokoloff_1994} Sokoloff, D., Nesme-Ribes, E.: 1994, A\&A 288, 293

\bibitem[1874]{sporer_1874} Sp\"{o}rer, G.: 1874, Publication der
Astronomischen Gesellschaft, XIII

\bibitem[1878]{sporer_1878} Sp\"{o}rer, G.: 1878, Publicationen des
Astrophysikalischen Observatoriums
zu Potsdam Nr. 1, Vol. 1, part 1, Wilhelm Engelmann, Leipzig

\bibitem[1880]{sporer_1880} Sp\"{o}rer, G.: 1880, ibid Nr. 5, Vol. 2,  part 1

\bibitem[1886]{sporer_1886} Sp\"{o}rer, G.: 1886, ibid Nr. 17, Vol. 4, part 4

\bibitem[1894]{sporer_1894} Sp\"{o}rer, G.: 1894, bid Nr. 32, Vol. 10, part 1

\bibitem[2007]{vaquero_2007} Vaquero, J.M.: 2007, Advances in Space Research 40, 929

\bibitem[2009]{usoskin_2009} Usoskin, I.G., Mursula, K., Arlt, R., Kovaltsov, G.A.: 2009, ApJ 700, L154

\bibitem[2003]{usoskin_2003} Usoskin, I.G., Mursula, K., Kovaltsov, G.A.: 2003, Sol. Phys. 218, 295

\bibitem[2006]{zolotova_2006} Zolotova, N.V.,  Ponyavin, D.I.: 2006, A\&A 449, L1

\bibitem[2007]{zolotova_2007} Zolotova, N.V.,  Ponyavin, D.I.: 2007, A\&A 470, L17

\bibitem[2009]{zolotova_2009} Zolotova, N.V., Ponyavin, D.I., Marwan, N., Kurths, J.: 2009, A\&A
503, 197




\end{thebibliography}
\end{document}